\documentclass{vietnam}

\def\be{\begin{equation}}
\def\ee{\end{equation}}
\def\bea{\begin{eqnarray}}
\def\eea{\end{eqnarray}}

\newcommand\lsim{\mathrel{\rlap{\lower4pt\hbox{\hskip1pt$\sim$}}
        \raise1pt\hbox{$<$}}}
\newcommand\gsim{\mathrel{\rlap{\lower4pt\hbox{\hskip1pt$\sim$}}
        \raise1pt\hbox{$>$}}}

\newcommand{\Photo}{}

\begin{document}
\vspace*{4cm}
\title{THE SUNYAEV-ZEL'DOVICH EFFECT AND LARGE-SCALE STRUCTURE}

\author{ J. C. HILL }

\address{Department of Astronomy, Pupin Hall, Columbia University, 550 W. 120${}^{th}$ St.,\\
New York, NY USA 10027}

\maketitle
\abstracts{
The thermal Sunyaev-Zel'dovich (tSZ) effect is the inverse-Compton scattering of cosmic microwave background (CMB) photons off hot, ionized electrons, primarily located in galaxy groups and clusters.  Recent years have seen immense improvement in our ability to probe cosmology and the astrophysics of the intracluster medium using the tSZ signal.  Here, I describe cross-correlations of the tSZ effect measured in \emph{Planck} data with gravitational lensing maps from \emph{Planck} and the Canada-France-Hawaii Telescope Lensing Survey, as well as hydrodynamical simulations which show that such measurements do not probe ``missing baryons,'' but rather the pressure of ionized gas in groups and clusters over a wide range of halo masses and redshifts.  I also present recent measurements of higher-order tSZ statistics using data from the Atacama Cosmology Telescope, which yield strong constraints on the amplitude of density fluctuations.  I describe stacking analyses of tSZ data from \emph{Planck}, focusing on the behavior of the gas pressure in low-mass galaxy groups.  I close with a prediction for the tSZ monopole, including relativistic corrections, which is the largest guaranteed spectral distortion signal for the proposed \emph{Primordial Inflation Explorer} mission.  The tSZ monopole will yield a direct measurement of the total thermal energy in ionized electrons in the observable universe.}

\section{Introduction}\label{sec:intro}

Analyses of the primary anisotropies in the cosmic microwave background (CMB) radiation have firmly established the standard model of cosmology.\cite{Hinshawetal2013,Planck2015params}  While the future of primordial CMB measurements lies in polarization, an abundance of information in the CMB temperature field remains to be extracted from the secondary anisotropies generated at redshift $z < 1100$.  Amongst these secondary sources is the thermal Sunyaev-Zel'dovich (tSZ) effect, the inverse-Compton scattering of CMB photons off hot, ionized electrons.\cite{Zel69,Sunyaev-Zeldovich1970}  The scattering generates a characteristic non-blackbody distortion in the spectrum of the CMB, leading to a decrement (increment) in the observed CMB temperature at frequencies below (above) $\approx 217$ GHz.  To lowest order, the amplitude of the ``Compton-$y$'' (tSZ) signal is proportional to the integrated electron pressure along the line-of-sight.

The majority of the electrons responsible for the tSZ signal are located in galaxy groups and clusters, where the virial temperature ($\sim 10^7$--$10^8$ K) and electron density ($\sim 0.001$--$0.01$ cm${}^{-3}$) are high enough to impose a measurable distortion on the CMB.  The Compton-$y$ field is thus strongly biased toward massive halos, especially in comparison to direct probes of the matter density field, such as gravitational lensing.  This bias can be an advantage --- for example, it is responsible for the strong sensitivity of tSZ statistics (e.g., the tSZ power spectrum or tSZ cluster counts) to some cosmological parameters, particularly $\sigma_8$ (the rms linear density fluctuation on the scale of 8 Mpc$/h$) and $\Omega_m$ (the matter density).  However, the bias renders it difficult to use the tSZ signal to probe baryons in cooler or more rarified states.  Instead, the tSZ signal is a sensitive probe of the thermal electron pressure profile of the intracluster medium (ICM).  Through the pressure profile, it is possible to look for deviations from self-similar ICM physics, as well as the signature of feedback energy injected by active galactic nuclei (AGN).  By extracting statistical measurements of the tSZ effect in CMB maps, constraints can simultaneously be placed on cosmological parameters and the population-level properties of the ICM as a function of mass and redshift.

I consider four such statistical approaches here.  In \S\ref{sec:tSZxlens}, I discuss cross-correlations of the tSZ signal with weak gravitational lensing maps, from both CMB and galaxy measurements.  In \S\ref{sec:tSZPDF}, I consider higher-order tSZ statistics beyond the power spectrum, focusing in particular on the one-point probability distribution function (PDF).  In \S\ref{sec:tSZstack}, I describe stacking analyses that constrain the tSZ signal as a function of galaxy stellar mass.  Finally, in \S\ref{sec:meany}, I present predictions for the mean tSZ signal of the universe, including both the standard non-relativistic $\langle y \rangle$ parameter and frequency-dependent relativistic corrections.

\section{Thermal SZ -- Gravitational Lensing Cross-Correlations}\label{sec:tSZxlens}

\begin{figure}
\begin{minipage}{1.0\linewidth}
\centerline{\includegraphics[width=0.65\linewidth]{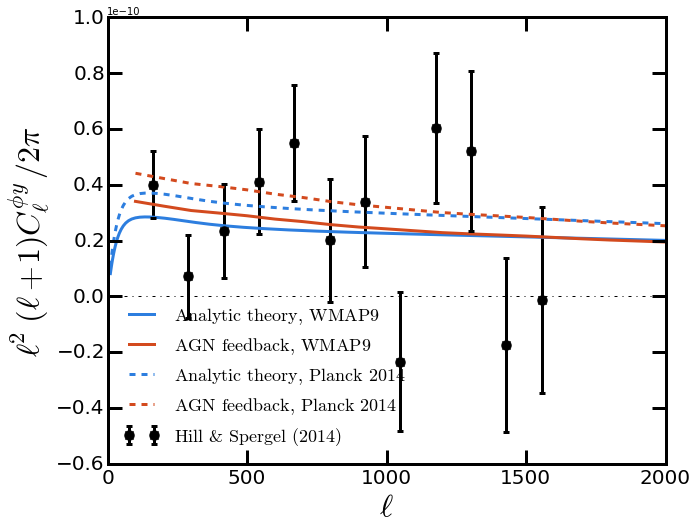}}
\end{minipage}
\caption[]{The tSZ ($y$) -- CMB lensing potential ($\phi$) cross-power spectrum.  (Original figure in Battaglia et al.\cite{BHM14} with data from Hill and Spergel.\cite{Hill-Spergel2014}) The black points show the cross-power spectrum measured from \emph{Planck} data.\cite{Hill-Spergel2014}  The red curves show the predicted signal from cosmological hydrodynamics simulations incorporating AGN feedback and other sub-grid physics,\cite{Battagliaetal2012} while the blue curves show an analytic halo model prediction computed self-consistently using a pressure profile model extracted from the simulations.  The small difference between the simulation and analytic curves at low-$\ell$ (large scales) is the contribution from diffuse, unbound gas.  The solid curves correspond to a WMAP9 cosmology ($\sigma_8 = 0.817$, $\Omega_m = 0.282$), while the dashed curves correspond to a \emph{Planck} 2013 cosmology ($\sigma_8 = 0.831$, $\Omega_m = 0.316$).  The data prefer a lower amplitude than that predicted by the \emph{Planck} cosmological parameters.}
\label{fig:BHM15Planck}
\end{figure}

Cross-correlations of the tSZ signal with gravitational lensing maps probe the relation between the pressure distribution of hot, ionized gas and the underlying (dark matter-dominated) density field.  The first detections of such cross-correlations were made in 2013 using data from \emph{Planck} and the Canada-France-Hawaii Telescope Lensing Survey (CFHTLenS).\cite{Hill-Spergel2014,vanWaerbekeetal2014}  Hill and Spergel\cite{Hill-Spergel2014} constructed a Compton-$y$ map from \emph{Planck} High Frequency Instrument (HFI) temperature maps at 100, 143, 217, 353, and 545 GHz by implementing a modified version of the constrained internal linear combination algorithm.\cite{Remazeillesetal2011}  Cross-correlating this $y$-map with the publicly released \emph{Planck} 2013 CMB lensing potential map\cite{Planck2013lensing} yielded a strong detection, but additional analysis was required to remove leakage of cosmic infrared background (CIB) emission into the $y$-map, as the CIB was already known to correlate strongly with CMB lensing.\cite{Planck2013CIBxlens}  The CIB leakage was assessed by cross-correlating the $y$-map with the \emph{Planck} 857 GHz map (a robust tracer of dust emission), and subtracting the appropriately weighted CIB -- CMB lensing correlation from the measured tSZ -- CMB lensing correlation.  The final result was a $6.2\sigma$ detection of the tSZ -- CMB lensing cross-power spectrum, shown in the black points in Fig.~\ref{fig:BHM15Planck}.  In the original analysis,\cite{Hill-Spergel2014} the measurement was interpreted with halo model calculations, using a variety of pressure profiles from the literature.\cite{Battagliaetal2012,Arnaudetal2010}  In the context of the ``universal pressure profile'' of Arnaud et al.,\cite{Arnaudetal2010} the results constrained the mean hydrostatic mass bias of the groups and clusters sourcing the signal to be $(1-b) = 1.06^{+0.11}_{-0.14}$, where $(1-b)$ is the ratio between the cluster mass inferred assuming hydrostatic equilibrium and the true cluster mass.  Note that this constraint includes groups and clusters over a wide mass ($10^{13} M_{\odot}/h \lsim M_{\rm virial} \lsim 10^{15} M_{\odot}/h$) and redshift ($0 \lsim z \lsim 2.5$) range, due to the high-redshift kernel probed by CMB lensing.

\begin{figure}
\begin{minipage}{1.0\linewidth}
\centerline{\includegraphics[width=0.55\linewidth]{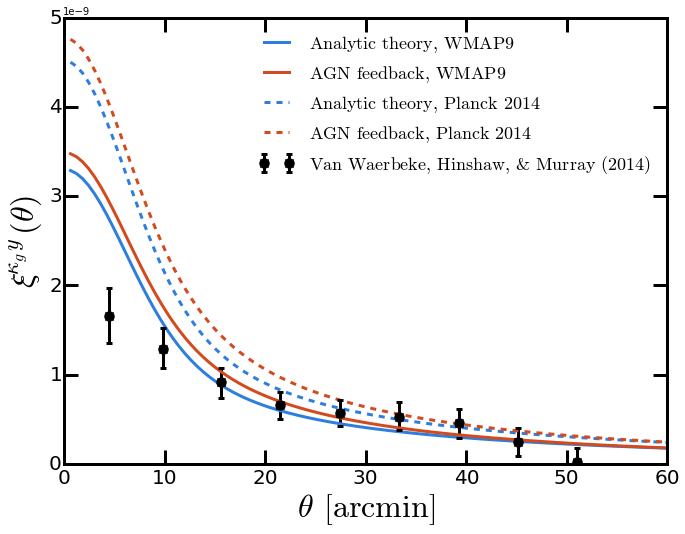}}
\end{minipage}
\caption[]{The tSZ ($y$) -- CFHTLenS lensing convergence ($\kappa_g$) cross-correlation function.  (Original figure in Battaglia et al.\cite{BHM14} with data from van Waerbeke et al.\cite{vanWaerbekeetal2014}) The black points show the cross-correlation measured using \emph{Planck} and CFHTLenS data.\cite{vanWaerbekeetal2014}  The theoretical curves follow the same conventions as in Fig.~\ref{fig:BHM15Planck}.  Note that the bin-to-bin correlations are strong (hindering a simple ``chi-by-eye'') because the measurement is in real space rather than harmonic space.  The analytic and simulation calculations agree closely on all scales, apart from slight deviations at small scales that could be due to baryonic effects on the halo density profiles.  In particular, there is no evidence for significant contributions from diffuse, unbound gas at large angular scales.  As in Fig.~\ref{fig:BHM15Planck}, the data prefer a lower amplitude than predicted by the \emph{Planck} cosmological parameters.}
\label{fig:BHM15CFHTLenS}
\end{figure}

A related signal was measured by van Waerbeke et al.,\cite{vanWaerbekeetal2014} who cross-correlated CFHTLenS weak gravitational lensing convergence data with Compton-$y$ maps constructed from \emph{Planck} HFI data at 100, 143, 217, and 353 GHz.  They formed a number of different $y$-maps constructed from different linear combinations of the \emph{Planck} data in order to assess residual contamination from non-tSZ sources.  Note that CIB contamination is less significant than in the tSZ -- CMB lensing measurement because the CFHTLenS galaxies are predominantly located at $z < 1$, whereas most of the CIB emission originates at higher redshifts.  The final result was a $6\sigma$ detection of the tSZ -- CFHTLenS cross-correlation function, shown in the black points in Fig.~\ref{fig:BHM15CFHTLenS} (note that this is a real-space cross-correlation, whereas Fig.~\ref{fig:BHM15Planck} is a harmonic-space cross-power spectrum).  In the original analysis,\cite{vanWaerbekeetal2014} the signal was interpreted with a constant gas bias model, in which the gas density is proportional to the mass density with a redshift-dependent bias factor, and fluctuations in the gas temperature are ignored (the temperature is assumed to evolve as $(1+z)^{-1}$).  The data were then interpreted to be the signal of warm ($T \sim 10^7$ K), diffuse ($n_e \sim 0.25$ m${}^{-3}$) baryons, rather than bound gas located in galaxy groups and clusters.

The physical origin of these signals was subsequently revisited\cite{BHM14} using the hydrodynamical simulations of Battaglia et al.\cite{Battagliaetal2010}, which allowed an assessment of the accuracy of the halo model approach used in the tSZ -- CMB lensing analysis and a quantification of the signal arising from bound and unbound gas (see also Hojjati et al.\cite{Hojjatietal2014}).  The comparison of the halo model and full simulation results for the tSZ -- CMB lensing signal is shown in Fig.~\ref{fig:BHM15Planck}, where the analytic calculations use an ICM pressure profile model extracted self-consistently from the same set of simulations.  Thus, any disagreement would point to a breakdown of the halo model assumptions (e.g., the assumption of an NFW density profile).  The agreement is excellent, except on large angular scales (low-$\ell$), where the simulations lie slightly higher than the halo model.  This excess is interpreted as the signal of diffuse, unbound gas (sometimes known as ``missing baryons''), which contributes $\approx 15$\% of the total signal at $\ell=500$ (eventually becoming negligible at high-$\ell$).  However, this contribution is degenerate with changes in cosmological and ICM pressure profile parameters, preventing a robust inference of its presence from the data.\cite{BHM14}  Combining the simulation results with parameter dependences derived from the halo model, Battaglia et al.\ used the data from Hill and Spergel to constrain $\sigma_8 \left( \Omega_m/0.282 \right)^{0.26} = 0.814 \pm 0.029$.

The halo model -- simulation comparison for the tSZ -- CFHTLenS signal is shown in Fig.~\ref{fig:BHM15CFHTLenS}, using the same approach as in Fig.~\ref{fig:BHM15Planck}, but presented as a real-space cross-correlation function in order to match the measurements (note that this choice yields strong bin-to-bin correlations).  In contrast to Fig.~\ref{fig:BHM15Planck}, the halo model matches the simulation results on nearly all scales, apart from minor differences at small scales that could result from baryonic effects in the inner regions of halo density profiles.  The close match on large angular scales implies that contributions from diffuse, unbound gas are negligible --- electrons located in halos can account for essentially the entire measured large-scale signal.  Without access to the full covariance matrix, cosmological parameters could not be fit, but it is clear that the data prefer an amplitude lower than that predicted by the \emph{Planck} cosmology.   A key takeaway from Battaglia et al.\cite{BHM14} is that both the tSZ -- CMB lensing and tSZ -- CFHTLenS cross-correlations, which probe the ICM over a vast range of masses and redshifts, are well-fit by a consistent pressure profile model and cosmology.  Forecasts for upcoming experiments indicate that ICM pressure profile parameters can be constrained to $\lsim 10$\% precision using such measurements.  At present, the small signal of diffuse, unbound gas (``missing baryons'') cannot be well-constrained using these methods, due to degeneracies with the ICM and cosmological parameters, but the techniques will prove extremely useful to constrain group and cluster pressure profiles in coming years.

\section{Higher-Order Thermal SZ Statistics}\label{sec:tSZPDF}

\begin{figure}
\begin{minipage}{1.0\linewidth}
\centerline{\includegraphics[width=0.7\linewidth]{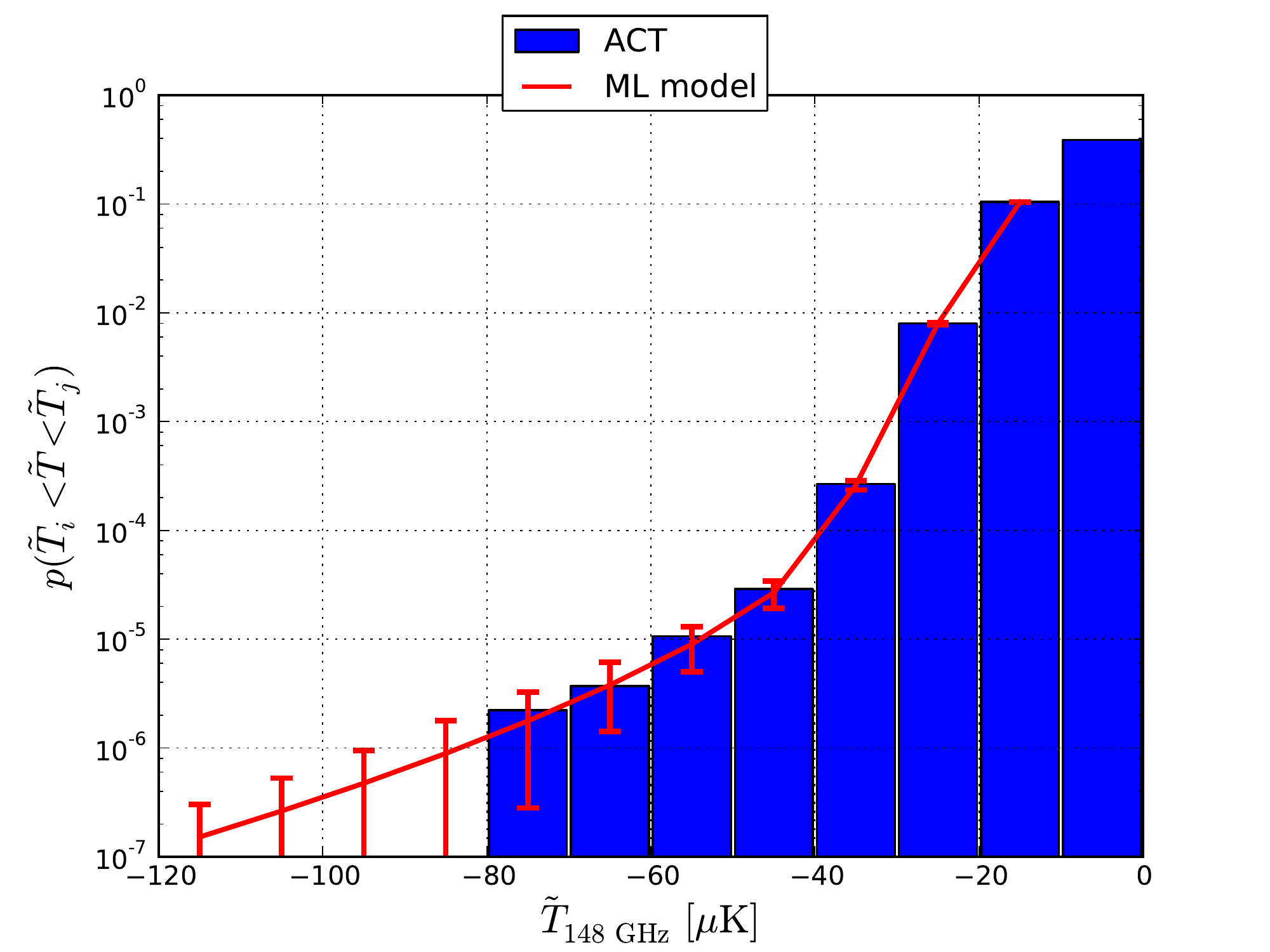}}
\end{minipage}
\caption[]{The tSZ one-point PDF measured in ACT data at 148 GHz.  (Original figure in Hill et al.\cite{Hilletal2014})  The blue bins show the measured histogram of Wiener-filtered ACT CMB temperature maps.  The red curve shows the maximum-likelihood model fit to these data, accounting for the tSZ signal and non-tSZ sources of noise (CMB, instrument/atmosphere, and foregrounds).}
\label{fig:ACTPDF}
\end{figure}

Statistical measures of the tSZ signal in CMB temperature maps, such as the tSZ power spectrum, have long been recognized as a powerful cosmological probe.\cite{Komatsu-Seljak2002,Hill-Pajer2013}  However, because the tSZ signal originates in rare, collapsed objects in the late-time density field, it is highly non-Gaussian, and thus the power spectrum does not capture all of the information contained in the tSZ field.  The first detection of a higher-order tSZ statistic, the tSZ real-space skewness, was made by the Atacama Cosmology Telescope (ACT) Collaboration\cite{Wilsonetal2012} in 2012 (the South Pole Telescope and \emph{Planck} Collaborations subsequently measured the tSZ bispectrum in 2013\cite{Crawfordetal2014,Planck2013ymap}).  The skewness signal was interpreted with simulations and halo model calculations, showing that it originated in more massive, lower redshift halos than those sourcing the tSZ power spectrum, and yielding a tight constraint on $\sigma_8$.  In a follow-up analysis, Hill and Sherwin\cite{Hill-Sherwin2013} demonstrated that different tSZ moments depend differently on underlying cosmological and ICM parameters, allowing degeneracies to be broken and tighter constraints to be obtained.

Hill et al.\cite{Hilletal2014} proceeded to consider the optimal extension of this idea: a measurement of the full tSZ one-point PDF, containing the information in all (zero-lag) moments of the field.  The PDF has the advantage of being a simple observable (a histogram), and using information from groups and clusters in the map that are below the threshold for individual detection.  Moreover, no modeling of the tSZ cluster selection function is required, as no individual objects are extracted from the map.  From a theoretical perspective, nearly all of the cosmological constraining power of the tSZ statistics is contained in their overall amplitudes, which depend sensitively on $\sigma_8$ due to their origin in the exponential tail of the halo mass function.  Very little cosmological constraining power is found in the shape of the tSZ power spectrum or bispectrum (although there is ICM parameter constraining power therein), and thus collapsing the information into the one-point PDF does not degrade cosmological constraints.  Finally, due to different effects on the shape of the tSZ PDF, it is possible to break degeneracies between cosmological and ICM parameters with a high signal-to-noise measurement.

Using ACT 148 GHz data, Hill et al.\cite{Hilletal2014} measured the tSZ PDF.  A theoretical approach was developed based on the halo model (and validated on simulations), with non-tSZ contributions from the CMB, foregrounds, and noise taken into account.  By focusing only on the negative temperature fluctuations at 148 GHz, nearly all non-tSZ foregrounds could be avoided, and any remainder was marginalized over.  The ICM model uncertainty was parametrized by an overall amplitude, which the data could not robustly constrain along with $\sigma_8$ and $\Omega_m$ (although future data will do so).  This parameter was thus marginalized over as well.  A small correction due to infrared sources ``filling in'' tSZ decrements at 148 GHz was applied based on 218 GHz measurements.  The final result was a tight constraint on cosmological parameters: $\sigma_8 \left(\Omega_m / 0.282\right)^{0.2} = 0.790 \pm 0.019\, ({\rm stat.})\, {}^{+0.018}_{-0.016} \, ({\rm ICM\,\, syst.})\, \pm 0.006\, ({\rm IR\,\, syst.})$.  The statistical error bar was nearly halved compared to the tSZ skewness analysis\cite{Wilsonetal2012} using essentially the same data, demonstrating the strong constraining power of higher tSZ moments.  Similar approaches were subsequently adopted in the 2015 Planck tSZ analysis.\cite{Planck2015ymap}

\section{Thermal SZ Stacking: Self-Similarity?}\label{sec:tSZstack}

With the advent of modern large-scale surveys, it is now possible to constrain the mean behavior of the hot, ionized gas in galaxy groups and clusters as a function of various observational proxies.  This approach is complementary to the statistical methods described above, as it explicitly relies on the identification of a sample of objects and external measurements of their (non-tSZ) properties.  In 2013, the \emph{Planck} Collaboration\cite{Planck2013LBGs} analyzed the stacked tSZ signal of a sample of ``locally brightest galaxies'' (LBGs) as a function of their stellar mass, $M_*$.  The galaxy sample was constructed from Sloan Digital Sky Survey Data Release 7, with various isolation criteria imposed to minimize satellite contamination and maximize the number of galaxies that were the central object in their dark matter halos.  The \emph{Planck} HFI temperature data were then used to extract the stacked tSZ signal of these objects ($Y$) as a function of their stellar mass, by means of a matched-filter approach.  The tSZ signal was detected at unprecedentedly low mass scales ($M_* \approx 10^{11.3} M_{\odot}$, corresponding to halo masses of order $5 \times 10^{12} M_{\odot}$).  Surprisingly, the results were consistent with a simple, self-similar, power-law $Y$--$M$ scaling relation, even at mass scales where the effects of feedback, gas depletion, and other non-thermal processes were expected to be important.  The effects of non-central galaxies, matched-filter miscentering, and other systematics were forward-modeled through the analysis using the Millennium simulation, from which ``effective'' halo masses corresponding to each of the stellar mass bins were derived.

The details of this analysis were subsequently revisited using both simulations\cite{LeBrunetal2015} and data.\cite{Grecoetal2015}  Several important caveats were identified in the follow-up analyses.  First, the tSZ signal was actually measured in a much larger physical aperture than presented in the \emph{Planck} results ($5 r_{500}$ rather than $r_{500}$, where $r_{500}$ refers to the radius enclosing a mean density 500 times the critical density at the halo redshift).  The $Y_{5r_{500}}$ values were then extrapolated to $Y_{500}$ values assuming a fixed template for the gas pressure profile, which was calibrated on much more massive objects and not known to be accurate for these halos.\cite{Arnaudetal2010}  The large aperture rendered the measured self-similar behavior less surprising, as even extreme AGN feedback models tend to preserve the cosmic mean gas fraction within $5r_{500}$, even for low-mass objects.\cite{LeBrunetal2015,Battagliaetal2013}  LeBrun et al.\cite{LeBrunetal2015} repeated the \emph{Planck} analysis using synthetic tSZ maps generated from simulations, and found that the assumption of an incorrect gas pressure profile template could bias the inferred $Y_{500}$ values high by up to an order of magnitude at the lowest mass scales probed.

\begin{figure}
\begin{minipage}{1.0\linewidth}
\centerline{\includegraphics[width=0.7\linewidth]{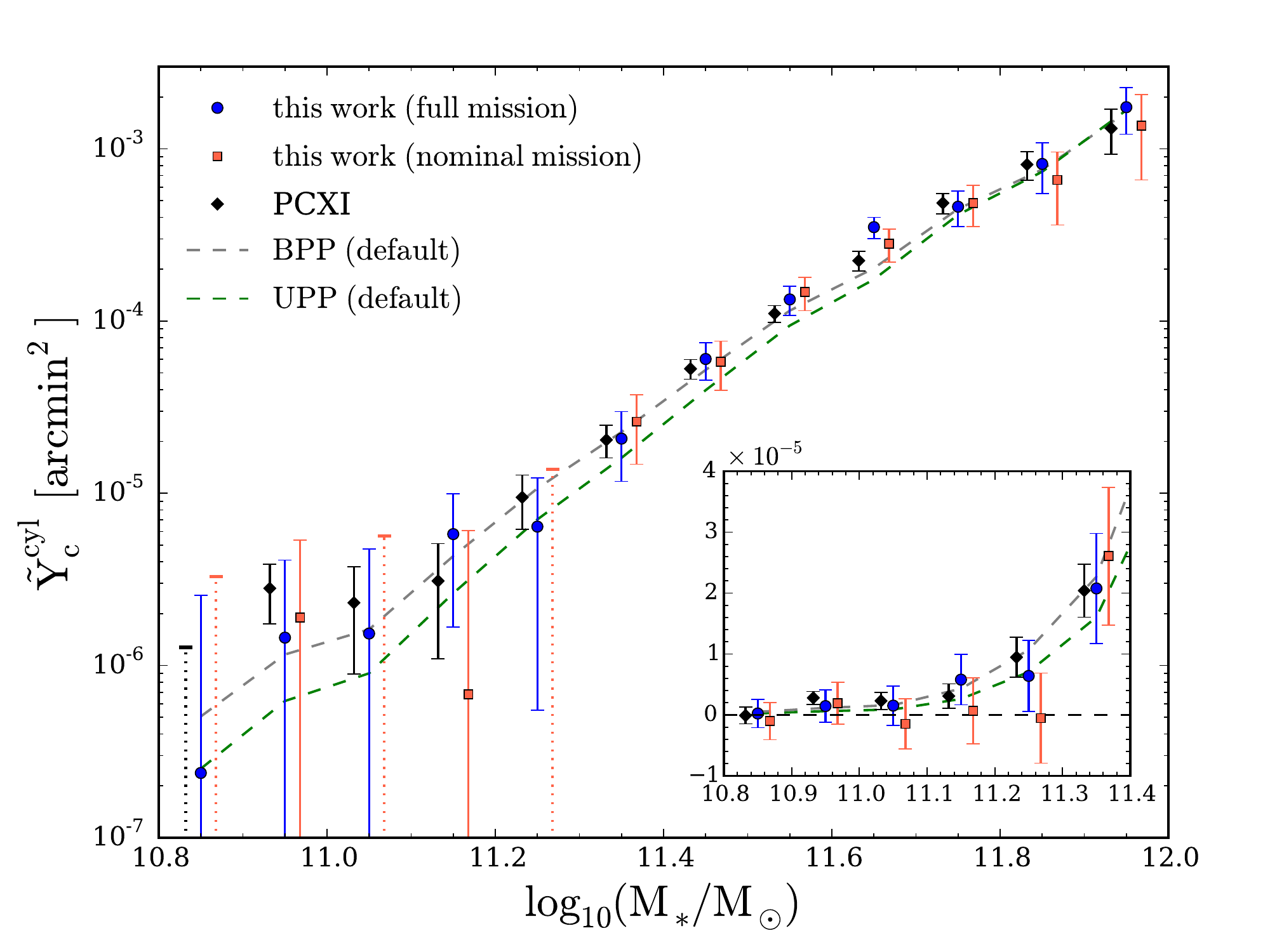}}
\end{minipage}
\caption[]{The stacked tSZ signal of ``locally brightest galaxies.''  (Original figure in Greco et al.\cite{Grecoetal2015})  The vertical axis shows the cylindrically integrated Compton-$y$ parameter within $5r_{500}$, while the horizontal axis shows the stellar mass of the galaxy.  The blue circles and red squares show the results of Greco et al.\cite{Grecoetal2015} using the \emph{Planck} full and nominal mission data, respectively.  Error bars are computed via bootstrap methods; the log-scale dotted error bars denote bins with a negative stacked signal (consistent with zero).  The black diamonds show the results of \emph{Planck} Collaboration XI.\cite{Planck2013LBGs}  The dashed black (green) curve shows the theoretical prediction of the Battaglia et al.\cite{Battagliaetal2012} (Arnaud et al.\cite{Arnaudetal2010}) pressure profile.}
\label{fig:Greco}
\end{figure}

Second, the \emph{Planck} results were presented in terms of unobserved quantities: the spherically integrated Compton-$y$ parameter (i.e., integrated within a sphere centered on each object), and the halo mass as inferred from an $M_*$--$M_{\rm halo}$ relation.  Motivated by these considerations and those discussed in the preceding paragraph, Greco et al.\cite{Grecoetal2015} revisited the \emph{Planck} LBG measurements using both the nominal and full \emph{Planck} mission data.  The stacked tSZ results are shown in Fig.~\ref{fig:Greco}, presented in terms of the stellar mass and the (observable) cylindrically integrated $Y$ parameter (within $5 r_{500}$ as in the \emph{Planck} analysis).  Instead of using a matched filter relying on an assumed template pressure profile, Greco et al. employed a simple aperture photometry technique, requiring minimal assumptions about the behavior of the gas.  In addition, they simultaneously modeled the tSZ and dust emission from each object in the sample, and found that the dust contamination was non-negligible (even at 100 GHz) in all LBGs except those of the highest masses in the sample.  Nonetheless, as shown in Fig.~\ref{fig:Greco}, the final results of the re-analysis were consistent with those from the \emph{Planck} study (after converting the \emph{Planck} results to the same observational plane).  Fig.~\ref{fig:Greco} also shows the predictions of two gas pressure profile models,\cite{Battagliaetal2012,Arnaudetal2010} which have been combined with the $M_*$--$M_{\rm halo}$ relation used in the \emph{Planck} analysis to yield results in this plane.  Note that a hydrostatic mass bias of $(1-b)=0.8$ has been assumed for the Arnaud et al.~profile.  Both models fit the data acceptably, with a slight preference for the Battaglia et al.~profile.  However, within the uncertainties, the data points are consistent with self-similar versions of these models, in which the integrated tSZ signal scales as $M^{5/3}$.  However, the results were found to be quite sensitive to the assumed $M_*$--$M_{\rm halo}$ relation, rendering a direct estimate of the $Y$-$M$ relation intractable.  

Note that the results in Fig.~\ref{fig:Greco} do not imply that the pressure profile obeys purely adiabatic gas physics, because the normalization of the $Y$-$M$ models shown does not correspond to this scenario.  For example, if the adiabatic model of Battaglia et al.\cite{Battagliaetal2012} were adopted instead, it would lie higher than the AGN feedback model (or Arnaud et al.~prescription) shown, even at the high-mass end.  Rather, the takeaway from this result is that the mass dependence of the integrated gas pressure is consistent with a self-similar ($M^{5/3}$) dependence, although the error bars are large enough that the small deviations from self-similarity predicted by the models are also consistent.  Upcoming measurements with higher resolution (allowing access to smaller scales than $5r_{500}$, where self-similar deviations should be larger) and lower noise levels will be needed to shed further light on the question of self-similarity.

\section{The Thermal SZ Monopole}\label{sec:meany}

Although the tSZ effect has now been measured in a vast number of individual objects, and its fluctuation properties probed through the power spectrum and other statistics, the mean tSZ signal imposed on the sky-averaged intensity spectrum of the CMB remains unmeasured.  The CMB intensity spectrum was last measured precisely by the \emph{COBE-FIRAS} experiment, which showed that the spectrum was consistent with that of a blackbody to $0.005$\% precision.\cite{Mather1994,Fixsen1996}  From this measurement, the mean Compton-$y$ parameter (i.e., mean non-relativistic tSZ signal) was constrained to be $|\langle y \rangle| < 1.5 \times 10^{-5}$ at 95\% confidence.\cite{Fixsen1996}  In recent years, interest has grown in new measurements of the CMB intensity spectrum,\cite{PIXIE2011} as spectral distortion data can potentially constrain a wide range of new physics, including energy injection from decaying or annihilating particles at early times,\cite{Hu-Silk1993} the dissipation of small-scale primordial density fluctuations (allowing constraints on the primordial power spectrum at very high wavenumber),\cite{Chluba2012} and energy injection from cosmic strings or primordial black holes.\cite{Ostriker-Thompson1987}

\begin{figure}
\begin{minipage}{0.5\linewidth}
\centerline{\includegraphics[width=0.9\linewidth]{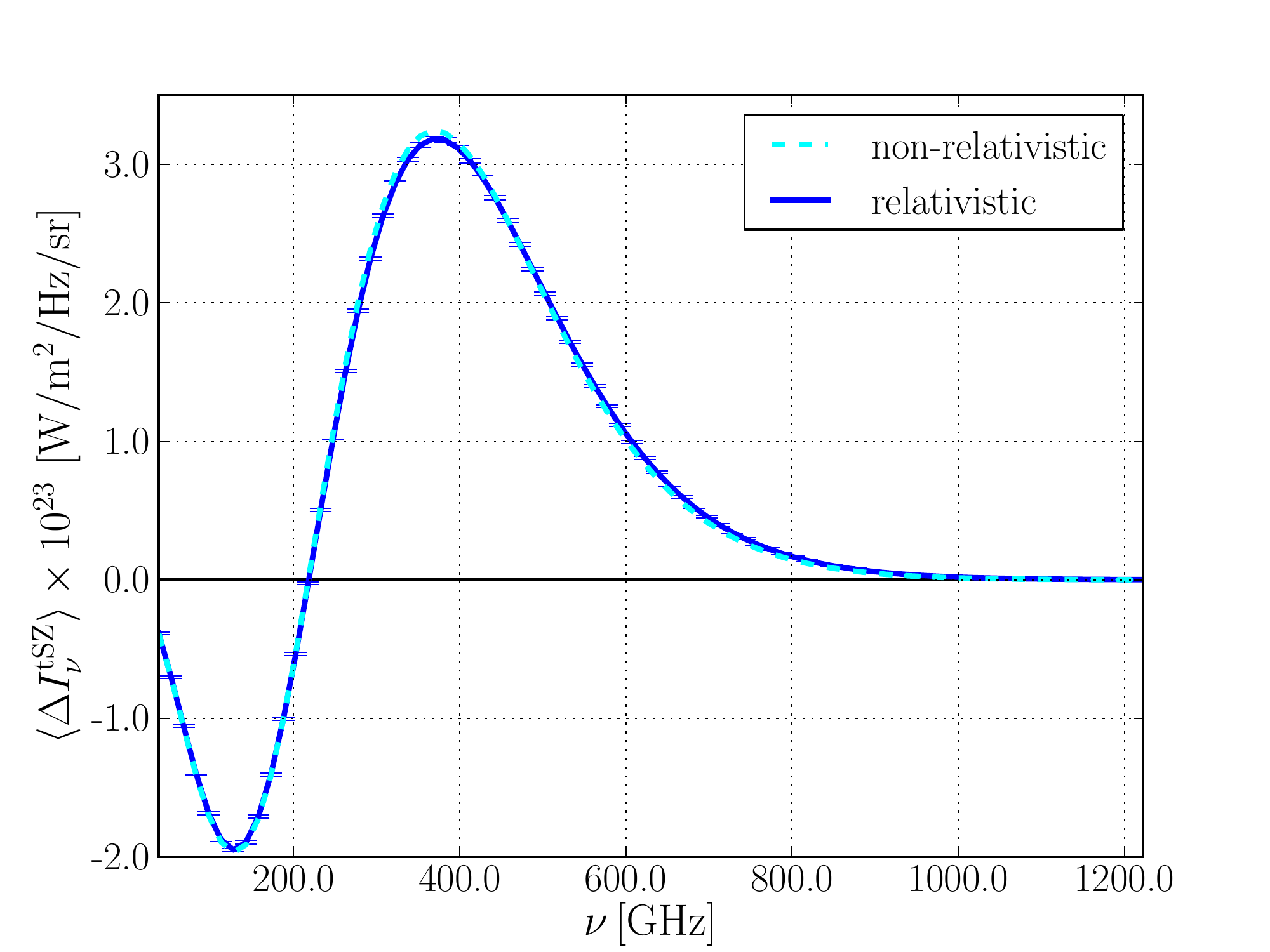}}
\end{minipage}
\hfill
\begin{minipage}{0.5\linewidth}
\centerline{\includegraphics[width=0.9\linewidth]{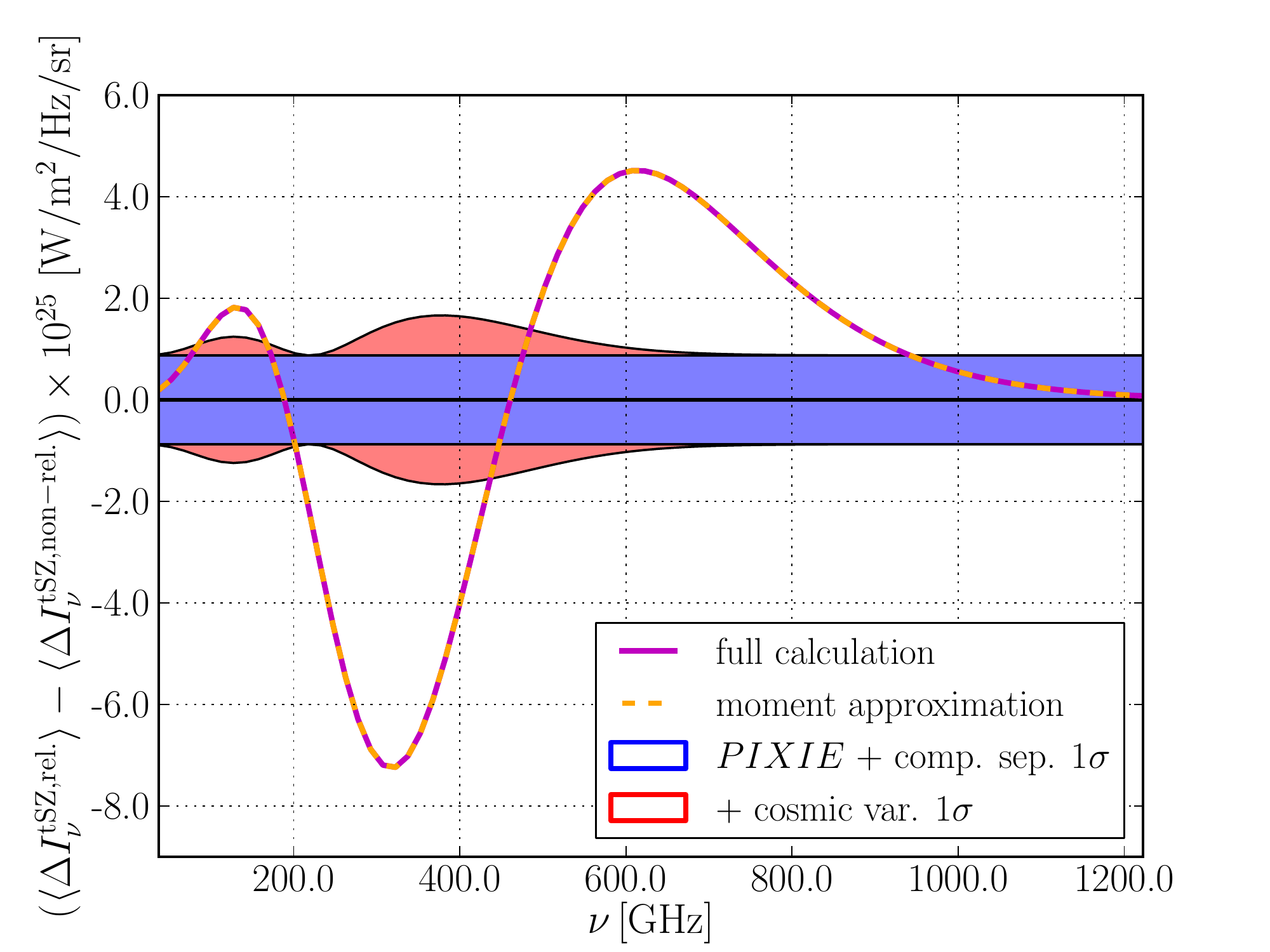}}
\end{minipage}
\caption[]{The predicted sky-averaged tSZ signal.  (Original figures in Hill et al.\cite{HBCFSS})  \emph{Left}: Non-relativistic (dashed cyan) and relativistic (solid blue) calculations.  The overall signal is dominated by hot, ionized electrons in galaxy groups and clusters, rather than the diffuse IGM or reionization.  Error bars for the proposed \emph{PIXIE} mission are shown on the relativistic curve, and include the \emph{PIXIE} noise, component separation noise, and cosmic variance.  \emph{Right}: Difference between the relativistic and non-relativistic predictions (solid magenta).  The dashed orange curve is an approximation based on moments of the optical depth-weighted ICM electron temperature distribution.  The shaded blue region is the \emph{PIXIE} noise and component separation noise, while the shaded red region is the cosmic variance uncertainty.}
\label{fig:meany}
\end{figure}

Amongst all of these potential sources of CMB spectral distortions, the largest guaranteed signal is that due to the tSZ effect.  Moreover, there is more to the tSZ signal than the non-relativistic $\langle y \rangle$.  Relativistic corrections\cite{Nozawaetal2006} arise from high-temperature electron populations ($k_B T_e \gsim 1$ keV), as found in massive halos.  Hill et al.\cite{HBCFSS} calculated the sky-averaged relativistic tSZ signal for the first time, using a halo model calibrated with a pressure profile from hydrodynamics simulations\cite{Battagliaetal2012} and a temperature--mass relation constrained by X-ray data.\cite{Arnaudetal2005}  In addition to the halo contributions, they also included contributions arising from the diffuse intergalactic medium (IGM) and reionization.  The total predicted Compton-$y$ parameter is roughly one order of magnitude below the \emph{COBE-FIRAS} bound (in agreement with early estimates\cite{Refregieretal2000}), $\langle y \rangle = 1.77 \times 10^{-6}$, and is dominated by the ICM contribution: $\langle y \rangle_{\rm ICM} = 1.58 \times 10^{-6}$, $\langle y \rangle_{\rm IGM} = 8.9 \times 10^{-8}$, and $\langle y \rangle_{\rm reion} = 9.8 \times 10^{-8}$.  The ICM signal is dominated by contributions from galaxy groups and low-mass clusters, primarily in the mass range $10^{12} M_{\odot}/h < M_{\rm virial} < 10^{14} M_{\odot}/h$.

The purely non-relativistic signal is compared to the full relativistic calculation in Fig.~\ref{fig:meany}.  The left panel shows the spectral distortion signal for the two cases, while the right panel shows the difference between the two.  In addition, the right panel shows an approximation to the full relativistic calculation based on moments of the optical depth-weighted ICM electron temperature distribution (following Chluba et al.\cite{ChlubaSwitzer2013}) --- the approximation is accurate to $\lsim 0.1$\%, and thus permits an interpretation of the signal in terms of these moments.  For example, the lowest-order moment is the mean optical depth-weighted ICM electron temperature.

In addition to these physical predictions, Fig.~\ref{fig:meany} also shows forecasted errors for the proposed \emph{Primordial Inflation Explorer} (\emph{PIXIE}), including the effects of foreground component separation, as well as the additional uncertainty from cosmic variance.  \emph{PIXIE} will detect the total signal at nearly $1500\sigma$ significance, although cosmic variance reduces the effective signal-to-noise to $230\sigma$.  Moreover, \emph{PIXIE} will detect relativistic corrections to the mean tSZ signal (i.e., the right panel of Fig.~\ref{fig:meany}) at $30\sigma$ significance.  These measurements will yield percent-level constraints on $\langle y \rangle$ and the mean optical depth-weighted ICM electron temperature, which will correspond to percent-level constraints on the total thermal energy in ionized electrons in the observable universe.  These measurements will impose a precise ``integral constraint'' on models of galaxy formation and feedback energy injection over cosmic time.  In addition, an understanding of this signal will be required in order to extract additional science from measurements of non-tSZ spectral distortions, such as the $\mu$ or ``residual'' (non-$y$/non-$\mu$) distortions.\cite{Chluba-Jeong2014}  Likewise, the tSZ signal from the ICM constitutes a ``foreground'' for possible measurements of the tSZ signal from reionization or the primordial universe, although it may be possible to subtract much of the ICM contribution through cross-correlations with deep galaxy and cluster catalogs.

%
%
\section*{Acknowledgments}
This work was supported by a Junior Fellow award from the Simons Foundation.  I also acknowledge support from NSF grant AST1311756.


\end{document}